\documentclass[review]{elsarticle}

\usepackage{lineno,hyperref}

\journal{Journal of The Franklin Institute}

\usepackage{tikz}

\usepackage{amsmath}
\usepackage{latexsym, amssymb}
\usepackage{graphicx}
\usepackage{amsmath, amsbsy}
\usepackage{amsopn, amstext}
\usepackage{hyperref}
\usepackage{cancel, color}
\usepackage{epstopdf}

\DeclareMathOperator{\sign}{sign}

\DeclareMathOperator{\Col}{Col}
\DeclareMathOperator{\Row}{Row}

\DeclareMathOperator{\GL}{GL}

\def\cal{\mathcal}

\def\ra{\rightarrow}

\def\a{\alpha}
\def\b{\beta}
\def\d{\delta}

\def\D{\Delta}

\def\0{{\bf 0}}

\newcommand{\R}{{\mathbb R}}

\def\dsum{\mathop{\sum}\limits}

\newtheorem{thm}{Theorem}
\newtheorem{dfn}[thm]{Definition}
\newtheorem{prp}[thm]{Proposition}
\newtheorem{exa}[thm]{Example}
\newtheorem{rem}[thm]{Remark}
\newtheorem{lem}[thm]{Lemma}

\newtheorem{cor}[thm]{Corollary}
\begin{document}

\begin{frontmatter}

\title{On Skew-Symmetric Games\tnoteref{mytitlenote}}
\tnotetext[mytitlenote]{This work is supported partly by NNSF
    61773371 and 61733018 of China.}


\author[mymainaddress]{Yaqi Hao\corref{mycorrespondingauthor}}
\cortext[mycorrespondingauthor]{Corresponding author}
\ead{hoayaqi@outlook.com}

\author[mysecondaryaddress]{Daizhan Cheng}
\address[mymainaddress]{School of Control Science and Engineering, Shandong University, Jinan 250061, P.R. China}
\address[mysecondaryaddress]{LSC, Academy of Mathematics and Systems Science,
Chinese Academy of Sciences, Beijing 100190, P.R.China}

\begin{abstract}
By resorting to the vector space structure of finite games, skew-symmetric games (SSGs) are proposed and investigated as a natural subspace of finite games. First of all, for two player games, it is shown that the skew-symmetric games form an orthogonal complement of the symmetric games. Then for a general SSG its linear representation is given, which can be used to verify whether a finite game is skew-symmetric. Furthermore, some properties of SSGs are also obtained in the light of its vector subspace structure. Finally, a symmetry-based decomposition of finite games is proposed, which consists of three mutually orthogonal subspaces: symmetric subspace, skew-symmetric subspace and asymmetric subspace. An illustrative example is presented to demonstrate this decomposition.
\end{abstract}


\end{frontmatter}


\section{Introduction}

The vector space structure of finite games is firstly proposed by \cite{can11}. Then it has been merged as an isomorphism onto a finite Euclidean space \cite{che16}.
As a result, the decomposition of the vector space of finite games becomes a natural and  interesting topic. Since the potential game is theoretically important and practically useful, a decomposition based on potential games and harmonic games has been investigated \cite{can11,che16}. Symmetric game is another kind of interesting games \cite{alo13}, which may provide useful properties for applications. Hence symmetry-based decomposition is another interesting topic. Decomposition may help to classify games and to reveal properties of each kind of finite games.

To provide a clear picture of the decompositions we first give a survey for the vector space structure of finite games.

\begin{dfn}\label{d1.1} A (normal form non-cooperative) finite game $G=(N,S,C)$ consists of three ingredients:
\begin{enumerate}
\item[(i)] Player, $N=\{1,2,\cdots,n\}$: which means there are $n$ players;
\item[(ii)] Profile, $S=\prod_{i=1}^nS_i$: where $S_i=\{1,\cdots,k_i\}$ is the set of strategies (actions) of player $i$;
\item[(iii)] Payoff, $C=\{c_1,c_2,\cdots,c_n\}$: where $c_i:S\ra \R$ is the payoff function of player $i$.
\end{enumerate}
\end{dfn}

Assume $j\in S_i$, i.e., $j$ is the $j$-th strategy of player $i$. Instead of $j$, we denote this strategy by $\d_{k_i}^j$, which is the $j$-th column of identity matrix $I_{k_i}$.
This expression is called the vector form of strategies. Since each payoff function $c_i$ is a pseudo-logical function, there is a unique row vector $V^c_i\in \R^k$, where $k=\prod_{i=1}^nk_i$, such that (when the vector form is adopted) the payoffs can be expressed as
\begin{align}\label{1.1}
c_i(x_1,\cdots,x_n)=V^c_i\ltimes_{j=1}^nx_j,\quad i=1,\cdots,n,
\end{align}
where $\ltimes$ is the semi-tensor product of matrices which is defined in next section.

The set of finite games $G=(N,S,C)$ with $|N|=n$, $|S_i|=k_i$, $i=1,\cdots,n$, is denoted by ${\cal G}_{[n;k_1,\cdots,k_n]}$. Now it is clear that a game $G\in  {\cal G}_{[n;k_1,\cdots,k_n]}$ is uniquely determined by
\begin{align}\label{1.2}
V_G:=\left[V^c_1,V^c_2,\cdots,V^c_n\right],
\end{align}
which is called the structure vector of $G$. Hence ${\cal G}_{[n;k_1,\cdots,k_n]}$ has a natural vector space structure as $\R^{nk}$.

The potential-based decomposition of finite games was firstly proposed by Candogan and Menache, using the knowledge of algebraic topology and the Helmholtz decomposition theory from graph theory \cite{can11}. The decomposition is shown in (\ref{1.3}), where ${\cal P}$, ${\cal N}$, and ${\cal H}$ are pure potential games, non-strategic games, and pure harmonic games respectively. Unfortunately, the inner product used there is not the standard one in $\R^{nk}$.
\begin{align}\label{1.3}
{\cal G}_{[n;k_1,\cdots,k_n]}=\rlap{$\underbrace{\phantom{\quad{\cal P}\quad\oplus\quad{\cal N}}}_{Potential\quad games}$}\quad{\cal P}\quad\oplus\quad
\overbrace{{\cal N}\quad\oplus\quad{\cal H}}^{Harmonic\quad games}.
\end{align}
The vector space structure of potential games has been clearly revealed in \cite{che14} by providing a basis of potential subspace. Using this result, \cite{che16} re-obtained
the decomposition (\ref{1.3}) with $\R^{nk}$ standard inner product through a  straightforward linear algebraic computation.

The concept of symmetric game was firstly proposed by Nash \cite{nas51}. It becomes an important topic since then \cite{alo13,bra09,cao16}. We also refer to \cite{chepr} for a vector space approach to symmetric games. The symmetry-based decompositions have been discussed recently as for four strategy matrix games \cite{sza15}, as well as for general two-player games \cite{sza16}.

In this paper, the skew-symmetric game is proposed. First, we show that two-player games have an orthogonal decomposition as in (\ref{1.4}). That is, the vector subspace of skew-symmetric games is the orthogonal complement of the subspace of symmetric games:
\begin{align}\label{1.4}
{\cal G}_{[2;\kappa]}={\cal S}_{[2;\kappa]}\oplus {\cal K}_{[2;\kappa]},
\end{align}
where ${\cal G}_{[2;\kappa]}={\cal G}_{[2;\kappa,\kappa]}$, ${\cal S}_{[2;\kappa]}$ and ${\cal K}_{[2;\kappa]}$ are symmetric and skew-symmetric subspaces of ${\cal G}_{[2;\kappa]}$ respectively.

Furthermore, certain properties of skew-symmetric games are also revealed. The bases of symmetric and skew-symmetric games are constructed. Due to their orthogonality, following conclusions about the decomposition of finite games ${\cal G}_{[n,\kappa]}$ are obtained:
\begin{itemize}
  \item [(1)] if $n~>~\kappa+1,$ then
  \begin{align}\label{1.401}
  {\cal G}_{[n;\kappa]}={\cal S}_{[n;\kappa]}\oplus {\cal E}_{[n;\kappa]};
  \end{align}
  \item [(2)] if $n~\leq~\kappa+1,$ then
  \begin{align}\label{1.402}
  {\cal G}_{[n;\kappa]}={\cal S}_{[n;\kappa]}\oplus {\cal K}_{[n;\kappa]}\oplus{\cal E}_{[n;\kappa]},
  \end{align}
where ${\cal E}_{[n;\kappa]}$ is the set of asymmetric games.
\end{itemize}

Finally, for statement ease, we give some notations:
\begin{enumerate}
\item  ${\cal M}_{m\times n}$: the set of $m\times n$ real matrices.

\item ${\cal B}_{m\times n}$: the set of $m\times n$ Boolean matrices, (${\cal B}_{n}$: the set of $n$ dimensional Boolean vectors.)

\item $\Col(M)$ ($\Row(M)$): the set of columns (rows) of $M$. $\Col_i(M)$ ($\Row_i(M)$): the $i$-th column (row) of $M$.

\item ${\cal D}:=\left\{0,1\right\}$.

\item $\d_n^i$: the $i$-th column of the identity matrix $I_n$.

\item $\D_n:=\left\{\d_n^i\vert i=1,\cdots,n\right\}$.

\item ${\bf 1}_{\ell}=(\underbrace{1,1,\cdots,1}_{\ell})^T$.

\item ${\bf 0}_{p\times q}$: a $p\times q$ matrix with zero entries.

\item A matrix $L\in {\cal M}_{m\times n}$ is called a logical matrix
if the columns of $L$ are of the form
$\d_m^k$. That is, $\Col(L)\subset \D_m$.
Denote by ${\cal L}_{m\times n}$ the set of $m\times n$ logical
matrices.

\item If $L\in {\cal L}_{n\times r}$, by definition it can be expressed as
$L=[\d_n^{i_1},\d_n^{i_2},\cdots,\d_n^{i_r}]$. For the sake of
compactness, it is briefly denoted as $
L=\d_n[i_1,i_2,\cdots,i_r]$.

\item ${\bf S}_n$: $n$-th order symmetric group.
\item $\left<\cdot,\cdot\right>$: the standard inner product in $\R^n$.
\item ${\bf P}_n$: $n$-th order Boolean orthogonal group.
\item $\GL(n,\R)$ (or~$\GL(V)$): general linear group.
\item ${\cal G}_{[n;k_1,\cdots,k_n]}$: the set of finite games with $|N|=n$, $|S_i|=k_i$.
\item ${\cal G}_{[n;\kappa]}$: $|S_i|=\kappa$, $i=1,\cdots,n$.
\item ${\cal S}_{[n;\kappa]}$: the set of (ordinary) symmetric games. Denote by $G_{S}$ a symmetric game.
\item ${\cal K}_{[n;\kappa]}$: the set of skew-symmetric games. Denote by $G_{K}$ a skew-symmetric game.
\item ${\cal E}_{[n;\kappa]}$: the set of asymmetric games. Denote by $G_{E}$ an asymmetric game.
\end{enumerate}

The rest of this paper is organized as follows: In section 2, a brief review of semi-tensor product of matrices is given. After introducing a symmetry-based classification of finite games, Section 3 presents mainly two results: (1) the orthogonal decomposition of two player games; (2) the linear representation of skew-symmetric games. Some properties of skew-symmetric games are discussed in Section 4. A basis of ${\cal K}_{[n,\kappa]}$ is also constructed. Section 5 is devoted to verifying the orthogonality of symmetric and skew-symmetric games. Section 6 provides a symmetry-based orthogonal decomposition of finite games. In Section 7, an illustrative example is given to demonstrate this decomposition. Section 8 is a brief conclusion.

\section{Preliminaries}

\subsection{Semi-tensor Product of Matrices}

In this section, we give a brief survey on semi-tensor product (STP) of matrices. It is the main tool for our approach. We refer to \cite{che11,che12} for details.
The STP of matrices is defined as follows:
\begin{dfn}Let $M \in {\cal M}_{m\times n},~~N \in {\cal M}_{p\times q}.$ The STP of $M$ and $N$ is defined as
\begin{align}\label{d.1}
M\ltimes N:= \left(M\otimes I_{t/n}\right)\left(N\otimes I_{t/p}\right)\in {\cal M}_{mt/n\times qt/p},
\end{align}
where $t=lcm(n,p)$ is the least common multiple of $n$ and $p,$ and $\otimes$ is the Kronecker product.
\end{dfn}

STP is a generalization of conventional matrix product, and all computational properties of the conventional matrix product remain available. It has been successfully used for studying logical (control) systems \cite{lht16,zgd17}. Throughout this paper, the default matrix product is STP. Hence, the product of two arbitrary matrices is well defined, and the symbol $\ltimes$ is mostly omitted.

First, we give some basic properties of STP, which will be used in the sequel.
\begin{prp} \label{pa.2}
 \begin{enumerate}
 \item (Associative Law:)
 \begin{align}\label{a.2}
 A\ltimes (B\ltimes C)=(A\ltimes B)\ltimes C.
 \end{align}
 \item (Distributive Law:)
 \begin{align}\label{a.3}
 (A+B)\ltimes C=A\ltimes C+B\ltimes C;
 \end{align}
 \begin{align}\label{a.4}
 C\ltimes (A+B)=C\ltimes A+C\ltimes B.
 \end{align}
 \end{enumerate}
\end{prp}
\begin{prp} \label{pa.3} Let $X\in \R^t$ be a $t$ dimensional column vector, and $M$ a matrix. Then
\begin{align}\label{a.5}
X\ltimes M=\left(I_t\otimes M\right)\ltimes X.
\end{align}
\end{prp}

\begin{dfn}\label{da.5} A swap matrix $W_{[m,n]}\in {\cal M}_{mn\times mn}$ is defined as
\begin{align}\label{a.7}
\begin{array}{ccl}
W_{[m,n]}&:=&[\d_n^1\d_m^1,\cdots,\d_n^n\d_m^1;\d_n^1\d_m^2,\cdots,\d_n^n\d_m^2,\cdots,
\d_n^1\d_m^m,\cdots,\d_n^n\d_m^m].
\end{array}
\end{align}
\end{dfn}

The basic function of a swap matrix is to swap two vectors.

\begin{prp}\label{pa.6} Let $X\in \R^m$ and $Y\in \R^n$ be two column vectors. Then
\begin{align}\label{a.8}
W_{[m,n]}XY=YX.
\end{align}
\end{prp}

The swap matrix is an orthogonal matrix:

\begin{prp}\label{pa.7} $W_{[m,n]}$ is an orthogonal matrix.  Precisely,
\begin{align}\label{a.9}
W^{-1}_{[m,n]}=W^{T}_{[m,n]}=W_{[n,m]}.
\end{align}
\end{prp}

Given a matrix $A=(a_{i,j})\in {\cal M}_{m\times n}$, its row stacking form is
$$
V_R(A):=\left(a_{11},a_{12},\cdots,a_{1n};\cdots;a_{m1},a_{m2},\cdots,a_{mn}\right)^T;
$$
its column stacking form is
$$
V_C(A):=\left(a_{11},a_{21},\cdots,a_{m1};\cdots;a_{1n},a_{2n},\cdots,a_{mn}\right)^T.
$$
Using Propositions \ref{pa.6} and \ref{pa.7} yields

\begin{prp}\label{pa.8} Given a matrix $A=(a_{i,j})\in {\cal M}_{m\times n}$.
Then
\begin{align}\label{a.10}
V_R(A)=W_{[m,n]}V_C(A);~V_C(A)=W_{[n,m]}V_R(A);
\end{align}
and
\begin{align}\label{a.11}
V_R(A^T)=V_C(A);\quad V_C(A^T)=V_R(A).
\end{align}
\end{prp}

Next, we consider the matrix expression of logical relations. Identifying
$$
1\sim \d_2^1,\quad 0\sim \d_2^2,
$$
then a logical variable $x\in {\cal D}$ can be expressed in vector form as
$$
x\sim\begin{pmatrix}x\\1-x\end{pmatrix},
$$
which is called the vector form expression of $x.$

A mapping $f:{\cal D}^n\ra \R$ is called a pseudo-Boolean function.

\begin{prp} \label{pa.4} Given a pseudo-Boolean function $f:{\cal D}^n\ra \R$,  there exists a unique row vector $V_f\in \R^{2^n}$, called the structure vector of $f$,  such that (in vector form)
\begin{align}\label{a.6}
f(x_1,\cdots,x_n)=V_f\ltimes_{i=1}^nx_i.
\end{align}
\end{prp}

\begin{rem}\label{ra.401} In previous proposition, if ${\cal D}$ is replaced by ${\cal D}_{\kappa}$, $\kappa>2$, then the function $f$ is called a pseudo-logical function and the expression (\ref{a.6}) remains available with an obvious modification that $x_i\in \D_{\kappa}$ and $V_f\in \R^{\kappa^n}$.
\end{rem}

\begin{dfn}\label{da.402} Let $A\in {\cal M}_{p\times n}$ and $B\in {\cal M}_{q\times n}$. Then the Khatri-Rao product of $A$ and $B$, denoted by $A*B$, is defined as follows:
$$\begin{array}{ccl}
A*B=[\Col_1(A)\ltimes \Col_1(B),\Col_2(A)\ltimes \Col_2(B),\cdots,\Col_n(A)\ltimes \Col_n(B)]\in {\cal M}_{pq\times n}.
\end{array}$$
\end{dfn}
\begin{prp} \label{pa.5} Assume $$u=M\ltimes_{i=1}^nx_{i}\in\D_p,~v=N\ltimes_{i=1}^nx_{i}\in\D_q,$$
where $x_{i}\in \D_{k_{i}},~i=1,2,\cdots,n;~M\in {\cal L}_{p\times k},~N\in {\cal L}_{q\times k},~k=\prod_{i=1}^nk_i.$ Then
$$uv=(M\ast N)\ltimes_{i=1}^nx_{i}\in\D_{pq}.$$
\end{prp}
\section{Symmetric and Skew-symmetric Games}
\subsection{Classification of Finite Games}
This subsection considers the symmetry-based classification of finite games.
First, we give a rigorous definition for symmetric and skew-symmetric games.

\begin{dfn}\label{d2.1} Let $G=(N,S,C)\in {\cal G}_{[n;\kappa]}$.
\begin{enumerate}
\item If for any $\sigma\in {\bf S}_n$, we have
\begin{equation}\label{2.1}
\begin{array}{l}
c_i(x_1,\cdots,x_n)=c_{\sigma(i)}(x_{\sigma^{-1}(1)},x_{\sigma^{-1}(2)},\cdots,x_{\sigma^{-1}(n)}),
\end{array}
\end{equation}
where $i=1,\cdots,n,$
then $G$ is called a symmetric game. Denote by ${\cal S}_{[n;\kappa]}$ the set of symmetric games in ${\cal G}_{[n;\kappa]}$.

\vspace{2mm}
\item
If for any $\sigma\in {\bf S}_n$, we have
\begin{equation}\label{2.2}
\begin{array}{l}
c_i(x_1,\cdots,x_n)=\sign(\sigma)c_{\sigma(i)}(x_{\sigma^{-1}(1)},x_{\sigma^{-1}(2)}\cdots, x_{\sigma^{-1}(n)}),
\end{array}
\end{equation}
where $i=1,\cdots,n,$
then $G$ is called a skew-symmetric game. Denote by ${\cal K}_{[n;\kappa]}$ the set of skew-symmetric games in ${\cal G}_{[n;\kappa]}$.
\end{enumerate}
\end{dfn}

It is well known that ${\cal G}_{[n;\kappa]}\sim \R^{n\kappa^n}$ is a vector space \cite{can11,che16}. It is easy to figure out that both ${\cal S}_{[n;\kappa]}$ and
${\cal K}_{[n;\kappa]}$ are subspaces of ${\cal G}_{[n;\kappa]}$. Hence, they are
also two subspaces of $\R^{n\kappa^{n}}$.

Then, we can define the following asymmetric subspace.
\begin{dfn}\label{d2.2}
$G$ is called an asymmetric game if its structure vector
\begin{align}\label{2.3}
V_G\in \left[{\cal S}_{[n;\kappa]} \bigcup {\cal K}_{[n;\kappa]}\right]^{\perp}.
\end{align}
The set of asymmetric games is denoted by ${\cal E}_{[n;\kappa]},$ which is also a subspace of $\R^{n\kappa^{n}}$.
\end{dfn}

\begin{exa}\label{e2.3}Consider $G\in {\cal G}_{[3;2]}.$ A straightforward computation shows the following result:
\begin{itemize}
\item [(1)]If $G\in {\cal S}_{[3;2]}$, then its payoff functions are as in Table \ref{Tab2.1}, where $[a,b,c,d,e,f]^{\mathrm{T}}\in \R^{6}.$

\item [(2)]If $G\in {\cal K}_{[3;2]},$ then its payoff functions are as in Table \ref{Tab2.2}, where $[g,h]^{\mathrm{T}}\in\R^{2}.$
\item [(3)]If $G\in {\cal E}_{[3;2]},$ then its payoff functions are as in Table \ref{Tab2.3}, where $[\xi_{1},\cdots,\xi_{16}]^{\mathrm{T}}\in\R^{16}$ and $\gamma_{1}=-\xi_{1}-\xi_{9},~\gamma_{2}=-\xi_{5}-\xi_{11},~\gamma_{3}=-\xi_{2}-\xi_{13}, ~\gamma_{4}=-\xi_{6}-\xi_{15},~\gamma_{5}=-\xi_{3}-\xi_{10},~\gamma_{6}=-\xi_{7}-\xi_{12},
~\gamma_{7}=-\xi_{4}-\xi_{14},~\gamma_{8}=-\xi_{8}-\xi_{16}.$
\end{itemize}

It follows that
$\dim(V_{{\cal S}_{[3,2]}})=6$, $\dim(V_{{\cal K}_{[3,2]}})=2$, and $\dim(V_{{\cal E}_{[3,2]}})=16$. Moreover, it is ready to verify the orthogonality:
$$\begin{array}{l}
V_{{\cal S}_{[3,2]}}V_{{\cal K}_{[3,2]}}^{T}=0,\\
V_{{\cal E}_{[3,2]}}V_{{\cal S}_{[3,2]}}^{T}=0,\\
V_{{\cal E}_{[3,2]}}V_{{\cal K}_{[3,2]}}^{T}=0.
\end{array}$$
We conclude that
$${\cal G}_{[3,2]}={\cal S}_{[3,2]} \oplus {\cal K}_{[3,2]}\oplus {\cal E}_{[3,2]},$$
which verifies (\ref{1.402}).

\begin{table}[!htbp] 
\centering \caption{Payoff matrix of a symmetric game in ${\cal S}_{[3;2]}$:\label{Tab2.1}}
\scalebox{0.8}{\begin{tabular}{|c|c|c|c|c|c|c|c|c|c|}
\hline
$ c\backslash a$ & 111 & 112 & 121 & 122 & 211 & 212 & 221 & 222 \\
\hline
$c_1$ & a & b & b & d & c & e & e & f\\
\hline
$c_2$ & a & b & c & e & b & d & e & f \\
\hline
$c_3$ & a & c & b & e & b & e & d & f \\
\hline
\end{tabular}}
\end{table}
\begin{table}[!htbp] 
\centering \caption{Payoff matrix of a skew-symmetric game in ${\cal K}_{[3;2]}$:\label{Tab2.2}}
\scalebox{0.8}{\begin{tabular}{|c|c|c|c|c|c|c|c|c|c|}
\hline
$ c\backslash s$ & 111 & 112 & 121 & 122 & 211 & 212 & 221 & 222 \\
\hline
$c_1$ & 0 & g & -g & 0 & 0 & h & -h & 0\\
\hline
$c_2$ & 0 & -g & 0 & -h & g & 0 & h & 0 \\
\hline
$c_3$ & 0 & 0& g & h & -g & -h & 0 & 0 \\
\hline
\end{tabular}}
\end{table}
\begin{table}[!htbp] 
\centering \caption{Payoff matrix of an asymmetric game in ${\cal E}_{[3;2]}$:\label{Tab2.3}}
\scalebox{0.8}{\begin{tabular}{|c|c|c|c|c|c|c|c|c|c|}
\hline
$ c\backslash s$ & 111 & 112 & 121 & 122 & 211 & 212 & 221 & 222 \\
\hline
$c_1$ & $\gamma_{1}$ & $\gamma_{2}$ & $\gamma_{3}$ & $\gamma_{4} $& $\gamma_{5} $& $\gamma_{6}$ & $\gamma_{7}$ & $\gamma_{8}$\\
\hline
$c_2$ & $\xi_{1}$ & $\xi_{2}$ & $\xi_{3}$ & $\xi_{4}$ & $\xi_{5}$ & $\xi_{6}$ & $\xi_{7}$ & $\xi_{8}$ \\
\hline
$c_3$ & $\xi_{9}$ & $\xi_{10}$ & $\xi_{11}$ & $\xi_{12}$ & $\xi_{13}$ & $\xi_{14}$ & $\xi_{15}$ & $\xi_{16}$ \\
\hline
\end{tabular}}
\end{table}
\end{exa}
\subsection{Two Player Games}

In this subsection we consider $G\in {\cal G}_{[2;\kappa]}$. Let $A$ and $B$ be the payoff matrices of player $1$ and player $2$ respectively. According to Definition \ref{d2.1}, it is easy to verify the following fact:
\begin{lem}\label{l20.1}
\begin{enumerate}
\item $G$ is a symmetric game, if and only if, $$A=B^T.$$
\item $G$ is a skew-symmetric game, if and only if, $$A=-B^T.$$
\end{enumerate}
\end{lem}

Note that for $G\in {\cal G}_{[2;\kappa]},$ we have its structure vector as
$$
V_G=[V_R^T(A),V_R^T(B)],
$$
where $V_R(A(B))$ is the row stacking form of matrix $A(B)$.

According to Propositions \ref{pa.7} and  \ref{pa.8}, we have the following result:

\begin{lem}\label{l20.2}
\begin{enumerate}
\item $G$ is a symmetric game, if and only if,
\begin{align}\label{20.1}
V_G=\left[V^c_1,V^c_1W_{[\kappa,\kappa]}\right].
\end{align}
\item $G$ is a skew-symmetric game, if and only if,
\begin{equation}\label{20.2}
V_G=\left[V^c_1,-V^c_1W_{[\kappa,\kappa]}\right].
\end{equation}
\end{enumerate}
\end{lem}


According to Lemma \ref{l20.2}, the following result can be obtained via a straightforward computation.
\begin{thm}\label{t20.3}
Let $G\in {\cal G}_{[2;\kappa]}$. Then $G$ can be orthogonally decomposed to
\begin{align}\label{20.3}
G=G_S\oplus G_K,
\end{align}
where $G_S\in {\cal S}_{[2;\kappa]}$ and $G_K\in {\cal K}_{[2;\kappa]}$.
\end{thm}
\noindent{\it Proof}. Denote the structure vector of $G$ as $V_G=[V^c_1,V^c_2]$.
We construct a symmetric game $G_S$ by setting
$$
V_{G_S}=[S,SW_{[\kappa,\kappa]}];
$$
and a skew-symmetric game $G_K$ by
$$
V_{G_K}=[K,-KW_{[\kappa,\kappa]}],
$$
where
$$
S=\frac{V^c_1+V^c_2W_{[\kappa,\kappa]}}{2};\quad K=\frac{V^c_1-V^c_2W_{[\kappa,\kappa]}}{2}.
$$
Then, it is ready to verify that
\begin{enumerate}
\item[(i)] $V_G=V_{G_S}+V_{G_K}$;

\vspace{2mm}
\item[(ii)] $\left<V_{G_S}, ~V_{G_K}\right>=0$.
\end{enumerate}
The conclusion follows.
\hfill $\Box$

Note that Theorem \ref{t20.3} implies the decomposition (\ref{1.4}).

\begin{exa}\label{e20.4} Consider $G\in {\cal G}_{[2;2]}$.
\begin{enumerate}
\item[(1)]
$G$ is symmetric, if and only if, its payoff functions are as in Table \ref{Tab20.1}.
\item[(2)] $G$ is skew-symmetric, if and only if, its payoff functions are as in Table \ref{Tab20.2}.
\item[(3)] Let $G\in {\cal G}_{[2;2]}$ with its payoff bi-matrix as in Table \ref{Tab20.3}.

Then it has an orthogonal decomposition into $G_S$ and $G_K$ with their payoff bi-matrices as in Table \ref{Tab20.1} and Table \ref{Tab20.2} respectively with
$$
\begin{array}{cccc}
a=\frac{\a+\b}{2},&b=\frac{\gamma+\eta}{2},&c=\frac{\xi+\d}{2}&\frac{\lambda+\mu}{2};\\
a'=\frac{\a-\b}{2},&b'=\frac{\gamma-\eta}{2},&c'=\frac{\xi-\d}{2}&\frac{\lambda-\mu}{2}.\\
\end{array}
$$
\begin{table}[h]
  \centering
  \caption{Payoff bi-matrix of a symmetric game in ${\cal G}_{[2;2]}$ \label{Tab20.1}}
  \scalebox{0.8}{\begin{tabular}{c|c|c}
   $P_1\backslash P_2$ & $1$ & $2$\\
   \hline
    $1$ & a,~a & b,~c\\
    \hline
    $2$ & c,~b & d,~d\\
  \end{tabular}}
\end{table}
\begin{table}[h]
  \centering
  \caption{Payoff bi-matrix of a skew-symmetric game in ${\cal G}_{[2;2]}$\label{Tab20.2}}
  \scalebox{0.8}{\begin{tabular}{c|c|c}
   $P_1\backslash P_2$ & $1$ & $2$\\
   \hline
    $1$ & a',~-a' & b',~-c'\\
    \hline
    $2$ & c',~-b' & d',~-d'\\
  \end{tabular}}
\end{table}
  \begin{table}[h]
  \centering
  \caption{Payoff bi-matrix of a game in ${\cal G}_{[2;2]}$ \label{Tab20.3}}
  \scalebox{0.8}{\begin{tabular}{c|c|c}
   $P_1\backslash P_2$ & $1$ & $2$\\
   \hline
    $1$ & $\a$,~$\b$ & $\gamma$,~$\d$\\
    \hline
    $2$ & $\xi$,~$\eta$ & $\lambda$,~$\mu$\\
  \end{tabular}}
\end{table}
\end{enumerate}
\end{exa}
\subsection{Skew-Symmetric Game and Its Linear Representation}
First, we present a necessary condition for verifying skew-symmetric games.

\begin{prp}\label{p3.1.1} Consider $G\in {\cal G}_{[n;\kappa]}$. If $G\in {\cal K}_{[n;\kappa]}$, then
\begin{align}\label{3.201}
V^c_i=-V^c_1W_{[\kappa^{i-2},\kappa]}W_{[\kappa, \kappa^{i-1}]},\quad i=2,\cdots,n.
\end{align}
\end{prp}
\noindent{\it Proof}. Consider $\sigma=(1,i)$. According to Definition \ref{d2.1}, we have
$$
c_{\sigma(1)}(x_{\sigma^{-1}(1)},\cdots,x_{\sigma^{-1}(n)})=-c_1(x_1,\cdots,x_n).
$$
That is,
$$
\begin{array}{l}
~~~V^c_ix_ix_2\cdots x_{i-1}x_1x_{i+1}\cdots x_n\\
=V^c_iW_{[\kappa^{i-1},\kappa]}x_2\cdots x_{i-1}x_1x_ix_{i+1}\cdots x_n\\
=V^c_iW_{[\kappa^{i-1},\kappa]}W_{[\kappa,\kappa^{i-2}]}x_1 x_2\cdots x_{i-1}x_ix_{i+1}\cdots x_n\\
=-V^c_1x_1x_2\cdots x_n.
\end{array}
$$
Hence, we have
$$
V^c_i W_{[\kappa^{i-1},\kappa]}W_{[\kappa,\kappa^{i-2}]}=
\sign(\sigma) V^c_1=-V^c_1.
$$
Then, (\ref{3.201}) follows from Proposition \ref{pa.7}.
\hfill $\Box$

Next, we consider another necessary condition: If $G\in {\cal K}_{[n;\kappa]}$, then what condition $c_1$ should verify?
An argument similar to the one used in Proposition \ref{p3.1.1} shows the following result.

\begin{prp}\label{p3.2} Consider $G\in {\cal G}_{[n;\kappa]}$. If $G\in {\cal K}_{[n;\kappa]}$, then
\begin{align}\label{3.102}
\begin{array}{l}
V^c_1\d_\kappa^s\left[I_{\kappa^{n-1}}+W_{[\kappa^{i-2},\kappa]}W_{[\kappa, \kappa^{i-3}]}\otimes I_{\kappa^{n-i}}\right]=0,
\end{array}
\end{align}
where $s=1,\cdots, \kappa;\;i=3,\cdots,n.$
\end{prp}
\noindent{\it Proof}. Assume $\sigma\in {\cal S}_{n}$ satisfies $\sigma(1)=1$. Let $\sigma=(2,i)$, $i>2$. Then, we have
$$
\begin{array}{l}
~~~~V^c_1x_1x_2\cdots x_n\\
=-V^c_1x_1x_{i}x_3\cdots x_{i-1}x_2x_{i+1}\cdots x_n\\
=-V^c_1x_1W_{[\kappa^{i-2},\kappa]}W_{[\kappa,\kappa^{i-3}]}x_{2}\cdots x_{n},\quad i=3,4,\cdots,n.
\end{array}
$$
Since $x_{2}\cdots x_{n}\in \D_{\kappa^{n-1}}$ are arbitrary, we have
$$V^c_1x_1=-V^c_1x_1W_{[\kappa^{i-2},\kappa]}W_{[\kappa,\kappa^{i-3}]}.$$
Setting $x_1=\d_{\kappa}^s$, we have (\ref{3.102}).
\hfill $\Box$

Note that the symmetric group ${\bf S}_n$ is generalized by transpositions $\{(1,i)\}$ \cite{chepr}. That is,
$$
{\bf S}_n=\left<(1,i)\;\vert\;1<i\leq n\right>.
$$
This fact motivates the following result.

\begin{thm}\label{t3.3} Consider $G\in {\cal G}_{[n;\kappa]}$.
\begin{enumerate}
\item[(1)] If $n=2$, then (\ref{3.201}) is the necessary and sufficient condition for $G\in {\cal K}_{[2;\kappa]}$.

\vspace{2mm}
\item[(2)] If $n>2$, then (\ref{3.201}) and (\ref{3.102}) are necessary and sufficient conditions for $G\in {\cal K}_{[n;\kappa]}$.
\end{enumerate}
\end{thm}

\noindent{\it Proof}. We need only to prove the sufficiency.
\begin{itemize}
  \item [(1)] if $n=2,$ (\ref{20.2}) implies the sufficiency.

\vspace{2mm}
  \item [(2)] if $n>2,$ we divide our proof into two steps.
\end{itemize}
First, we prove the condition for a single payoff function $c_{i}$. For any $\sigma\in {\bf S}_n$ and $\sigma(1)=1,$ without loss of generality, we assume
\begin{equation}
\sigma:=(2,i_{1})(2,i_{2})\cdots(2,i_{t}):=\sigma_{1}\circ\sigma_{2}\circ\cdots\circ\sigma_{t},\nonumber
\end{equation}
where $\sigma_{j}=(2,i_{j}),~j=1,2,\cdots,t.$

From (\ref{3.102}), it can be calculated that
\begin{equation}\label{11.11}
\begin{array}{l}
~~~V^c_1x_1x_2\cdots x_n\\
=-V^c_1x_1x_{\sigma_{t}^{-1}(2)}\cdots x_{\sigma_{t}^{-1}(n)}\\
=(-1)^{2}V^c_1x_1x_{\sigma_{t}^{-1}(\sigma_{t-1}^{-1}(2))}\cdots x_{\sigma_{t}^{-1}(\sigma_{t-1}^{-1}(n))}\\
\vdots\\
=(-1)^{t}V^c_1x_1x_{\sigma_{t}^{-1}(\cdots(\sigma_{1}^{-1}(2)))}\cdots x_{\sigma_{t}^{-1}(\cdots(\sigma_{1}^{-1}(n)))}\\
=sgn(\sigma)V^c_1x_1x_{\sigma^{-1}(1)}\cdots x_{\sigma^{-1}(n)}
\end{array}
\end{equation}
Applying (\ref{3.201}) to (\ref{11.11}), we have
\begin{align}\label{11.12}
\begin{array}{l}
~~~~V^c_ix_{i}x_{2}\cdots x_{i-1} x_{1}x_{i+1}\cdots x_{n}\\
=sgn(\sigma)V^c_ix_{\sigma^{-1}(i)}x_{\sigma^{-1}(2)}\cdots x_{\sigma^{-1}(i-1)}x_{1}x_{\sigma^{-1}(i+1)}\cdots x_{\sigma^{-1}(n)}.
\end{array}
\end{align}
(\ref{11.12}) implies that for any $\sigma\in {\bf S}_n$ and $\sigma(i)=i,$ we have
\begin{equation}\label{11.10}
\begin{array}{l}
c_{i}(x_{1},\cdots,x_{i-1},x_{i},x_{i+1},\cdots,x_{n})=c_{i}(x_{\sigma^{-1}(1)},\cdots,x_{\sigma^{-1}(i-1)},x_{\sigma^{-1}(i)},\\
\quad\quad\quad\quad\quad\quad\quad\quad\quad\quad\quad\quad\quad\quad\quad\quad \quad\quad\quad\quad x_{\sigma^{-1}(i+1)},\cdots,x_{\sigma^{-1}(n)}).
\end{array}
\end{equation}
Obviously, according to (\ref{2.2}), (\ref{11.10}) is the necessary and sufficient condition for a single payoff function to obey in a skew-symmetric game.

Next, we consider the condition for cross payoffs.

For any $\sigma\in {\bf S}_n,$ without loss of generality, we assume
$$\sigma:=(1,i_{1})(1,i_{2})\cdots (1,i_{t}):=\sigma_{1}\cdots\sigma_{t},$$
where $\sigma_{j}=(1,i_{j}),~j=1,2,\cdots,t.$

Combining (\ref{3.201}) with (\ref{3.102})  yields
\begin{equation}\label{11.13}
\begin{array}{l}
~~~~V_{i}^{c}x_{1}\cdots x_{i-1}x_{i}x_{i+1}\cdots x_{n}\\
=-V_{\sigma_{t}(i)}^{c}x_{\sigma_{t}^{-1}(1)}x_{\sigma_{t}^{-1}(2)}\cdots x_{\sigma_{t}^{-1}(n)}\\
=(-1)^{2}V_{\sigma_{t-1}(\sigma_{t}(i))}^{c}x_{\sigma_{t}^{-1}(\sigma_{t-1}(1))}\cdots x_{\sigma_{t}^{-1}(\sigma_{t-1}(n))}\\
\vdots\\
=(-1)^{t}V_{\sigma_{1}(\cdots(\sigma_{t}(i)))}^{c}x_{\sigma_{t}^{-1}(\cdots(\sigma_{1}^{-1}(1)))}\cdots x_{\sigma_{t}^{-1}(\cdots(\sigma_{1}^{-1}(n)))}\\
=sgn(\sigma)V_{\sigma(i)}^{c}x_{\sigma^{-1}(1)}x_{\sigma^{-1}(2)}\cdots x_{\sigma^{-1}(n)}
\end{array}
\end{equation}
Clearly, (\ref{2.2}) follows from (\ref{11.10}) and (\ref{11.13}).
\hfill $\Box$

\begin{rem}\label{r3.301} When $n=2$ (\ref{3.201}) degenerated
\begin{align}\label{p3.401}
V^c_2=-V^c_1W_{[\kappa,\kappa]},
\end{align}
which coincides with (\ref{20.2}).
\end{rem}

\begin{exa}\label{e3.4}Consider $G\in {\cal{K}}_{[3;3]}.$ From Proposition \ref{p3.2} we have
\begin{equation}
V^c_1\d_3^s\left[I_{3^{2}}+W_{[3,3]}\right]=0,~s=1,2,3.\nonumber
\end{equation}
One can easily figure out that
$$\begin{array}{l}
V^c_1=[0,a_{1},a_{2},-a_{1},0,a_{3},-a_{2},-a_{3},0,0,b_{1},b_{2},-b_{1},0,b_{3},-b_{2},-b_{3},0,0,c_{1},c_{2},-c_{1},0,c_{3},-c_{2},-c_{3},0],
\end{array}$$
where $a_{i},b_{i},c_{i},~i=1,2,3,$ are real numbers.
According to Proposition \ref{p3.1.1}, we can calculate that
$$\begin{array}{l}
V^c_2=[0,-a_{1},-a_{2},0,0,-b_{1},-b_{2},-c_{1},-c_{2},a_{1},0,-a_{3},b_{1},0,-b_{3},c_{1},0,-c_{3},a_{2},a_{3},0,b_{2},b_{3},0,c_{2},c_{3},0],\\
V^c_3=[0,0,0,a_{1},b_{1},c_{1},a_{2},b_{2},c_{2},-a_{1},-b_{1},-c_{1},0,0,0,a_{3},b_{3},c_{3},-a_{2},-b_{3},-c_{2},-a_{3},-b_{3},-c_{3},0,0,0].
\end{array}$$
According to Definition \ref{d2.1}, a straightforward verification shows that $G\in {\cal{K}}_{[3;3]}.$

As a byproduct, we have $\dim({\cal{K}}_{[3;3]})=9$.
\end{exa}

In the following, we consider the linear representation of ${\bf S}_{n}$ in ${\cal G}_{[n;\kappa]}.$

\begin{dfn}\label{d3.201} \cite{ser77} Let $A$ be a group and $V$ a finite dimensional vector space. A linear representation of $A$ in $V$ is a group homomorphism $\varphi: A \ra GL(V)$.
\end{dfn}

Consider a profile $s=(i_1,i_2,\cdots,i_n)$ of a $G\in {\cal G}_{[n;\kappa]}$. We define two expressions of $s$ as follows:

\begin{itemize}
\item STP Form: The STP form of $s$ is expressed as
$$
s=\ltimes_{j=1}^n\d_{\kappa}^{i_j}.
$$
\item Stacking Form: The strategy stacking form of $s$ is expressed as
$$
\vec{s}=\begin{bmatrix}
\d_{\kappa}^{i_1}\\
\d_{\kappa}^{i_2}\\
\vdots\\
\d_{\kappa}^{i_n}\\
\end{bmatrix}.
$$
\end{itemize}

Denote
$$
\Phi:=\begin{bmatrix}
\Phi_1\\
\Phi_2\\
\vdots\\
\Phi_n
\end{bmatrix},
$$
where
$$
\Phi_{i}={\bf 1}_{\kappa^{i-1}}^{T}\otimes I_{k}\otimes{\bf 1}_{\kappa^{n-i}}^{T},\quad i=1,\cdots,n.
$$
It is easy to verify that the $\Phi$ can convert a profile from its STP form to its strategy stacking form. Precisely, we have
\begin{align}\label{3.40100}
\vec{s}=\Phi s.
\end{align}

\begin{rem}\label{r3.40100}
In the pseudo-logical function expression of payoffs (\ref{1.1}), the profiles are expressed in its STP form, while in Definition \ref{d2.1} to permute the strategies the stacking form of $s$ is convenient. That is why the above conversion is necessary.
\end{rem}

Using above notations and Definition \ref{d2.1}, the following result can be obtained easily.

\begin{prp}\label{p3.5} A game $G\in {\cal G}_{[n;\kappa]}$ is skew-symmetric, if and only if,
\begin{align}\label{3.5}
V_{i}^{c}=sgn(\sigma)V_{\sigma(i)}^{c}T_{\sigma},~\forall \sigma\in {\bf S}_{n},~i=1,\cdots,n,
\end{align}
where $T_{\sigma}=\Phi_{\sigma^{-1}(1)}\ast\Phi_{\sigma^{-1}(2)}\ast\cdots\Phi_{\sigma^{-1}(n)}$, and $\ast$ is the Khatri-Rao product.
\end{prp}

We need the matrix expression of a $\sigma\in {\bf S}_n$, denoted by $P_{\sigma}$. It is defined as
$$
\Col_i(P_{\sigma})=\d_n^j,\quad \mbox{if}~ \sigma(i)=j,\quad i=1,\cdots,n.
$$
Consequently, we have
\begin{align}\label{11.5}
P_{\sigma}=[\d_n^{\sigma(1)},\d_n^{\sigma(2)},\cdots,\d_n^{\sigma(n)}].
\end{align}
The next property follows immediately from the construction.

\begin{prp}\label{p3.501} Let $\sigma\in {\bf S}_n$. Then $\sigma(i)=j$, if and only if,
$$
P_{\sigma}\d_n^i=\d_n^j.
$$
\end{prp}
Denote $$V_{G}^{\sigma}=[V_{\sigma(1)}^{c},V_{\sigma(2)}^{c},\cdots,V_{\sigma(n)}^{c}]\in R^{n\kappa^{n}}.$$
Using Propositions \ref{p3.5} and \ref{p3.501}, we have the following theorem.

\begin{thm}\label{t3.4}$G\in {\cal G}_{[n;\kappa]}$ is skew-symmetric, if and only if,
\begin{align}\label{3.6}
V_{G}=V_{G}(P_{\sigma}\otimes sgn(\sigma)T_{\sigma}),~~\forall \sigma\in {\bf S}_{n}.
\end{align}
\end{thm}
\noindent{\it Proof}. (Necessity) From (\ref{3.5}) we have
\begin{eqnarray}
V_{G}&=&[V_{1}^{c},V_{2}^{c},\cdots,V_{n}^{c}]\nonumber\\
&=&[sgn(\sigma)V_{\sigma(1)}^{c}T_{\sigma},sgn(\sigma)V_{\sigma(2)}^{c}T_{\sigma},\cdots,sgn(\sigma)V_{\sigma(n)}^{c}T_{\sigma}]\nonumber\\
&=&[V_{\sigma(1)}^{c},V_{\sigma(2)}^{c},\cdots,V_{\sigma(n)}^{c}][I_{n}\otimes sgn(\sigma)T_{\sigma}]\nonumber\\
&=&V_{G}^{\sigma}[I_{n}\otimes sgn(\sigma)T_{\sigma}]\nonumber\\
&=&V_{G}P_{\sigma}[I_{n}\otimes sgn(\sigma)T_{\sigma}]\nonumber\\
&=&V_{G}[P_{\sigma}\otimes sgn(\sigma)T_{\sigma}].\nonumber
\end{eqnarray}

(Sufficiency) Splitting (\ref{3.6}) into blocks, where each block corresponds to a component $V_{i}^{c}$, a straightforward computation verifies (\ref{3.5}) .
\hfill $\Box$

\begin{prp}\label{p3.6}Define a mapping of $\sigma\in {\bf S}_{n}$ to the general linear group of the vector space $G\in {\cal G}_{[n;\kappa]}$ as
\begin{align}\label{3.7}
\psi(\sigma):=P_{\sigma}\otimes sgn(\sigma)T_{\sigma}\in GL({\cal G}_{[n;\kappa]}).
\end{align}
Then, $\psi$ is a linear representation of ${\bf S}_{n}$ in ${\cal G}_{[n;\kappa]}.$
\end{prp}

\noindent{\it Proof}. We need only to show that $\psi$ is a group homomorphism, that is
$$\psi(\mu\circ\sigma)=\psi(\mu)\psi(\sigma).$$
First, we verify that
\begin{align}\label{11.6}
P_{\mu\circ\sigma}=P_{\mu}P_{\sigma}.
\end{align}
Denote $$\textbf{P}_{n}:=\{P_{\sigma}\mid\sigma\in S_{n}\}.$$
Clearly, there is a one-one correspondence between $\textbf{P}_{n}$ and ${\bf S}_{n}.$ Furthermore, from (\ref{11.5}) we can calculate that
\begin{equation}
\begin{array}{ccl}
P_{\mu\circ\sigma}&=&[\d_n^{(\mu\circ\sigma)(1)},\d_n^{(\mu\circ\sigma)(2)},\cdots,\d_n^{(\mu\circ\sigma)(n)}]\nonumber\\
&=&[\d_n^{\mu(\sigma(1))},\d_n^{\mu(\sigma(2))},\cdots,\d_n^{\mu(\sigma(n))}]\nonumber\\
&=&P_{\mu}[\d_n^{\sigma(1)},\d_n^{\sigma(2)},\cdots,\d_n^{\sigma(n)}]\nonumber\\
&=&P_{\mu}P_{\sigma}.\nonumber
\end{array}
\end{equation}
Next, we prove that
\begin{align}\label{11.7}
T_{\mu\circ\sigma}=T_{\mu}T_{\sigma}.
\end{align}
Define
$$\sigma(x)=\ltimes_{i=1}^nx_{\sigma^{-1}(i)}.$$
According to Proposition \ref{pa.5} and (\ref{3.40100}) we have
\begin{equation}
\begin{array}{ccl}
\sigma(x)&=&\ltimes_{i=1}^n (\Phi_{\sigma^{-1}(i)}x)\nonumber\\
&=&(\Phi_{\sigma^{-1}(1)}x)(\Phi_{\sigma^{-1}(2)}x)\cdots(\Phi_{\sigma^{-1}(n)}x)\nonumber\\
&=&(\Phi_{\sigma^{-1}(1)}\ast\Phi_{\sigma^{-2}(2)}\ast\cdots\ast\Phi_{\sigma^{-1}(n)})x\nonumber\\
&=&T_{\sigma}x. \nonumber
\end{array}
\end{equation}
Hence $$T_{\mu\circ\sigma}(x)=(\mu\circ\sigma)(x)=\mu(\sigma(x))=\mu(T_{\sigma}x)=T_{\mu}T_{\sigma}x.$$
Note that
\begin{equation}\label{11.8}
sgn(\mu\circ\sigma)=sgn(\mu)sgn(\sigma).
\end{equation}
Using (\ref{11.6}), (\ref{11.7}) and (\ref{11.8}), a straightforward computation shows that
\begin{equation}
\begin{array}{ccl}
\psi(\mu\circ\sigma)&:=&P_{\mu\circ\sigma}\otimes sgn(\mu\circ\sigma)T_{\mu\circ\sigma}\nonumber\\
&=&(P_{\mu}P_{\sigma})\otimes[sgn(\mu)sgn(\sigma)T_{\mu}T_{\sigma}]\nonumber\\
&=&[P_{\mu}\otimes sgn(\mu)T_{\mu}][P_{\sigma}\otimes sgn(\sigma)T_{\sigma}]\nonumber\\
&=&\psi(\mu)\psi(\sigma).\nonumber
\end{array}
\end{equation}
The proof is completed.
\hfill $\Box$

Theorem \ref{t3.4} and Proposition \ref{p3.6} lead to the following result.
\begin{cor}\label{c3.7}$G\in {\cal G}_{[n;\kappa]}$ is skew-symmetric, if and only if, it is invariant with respect to the linear representation $\psi(\sigma),~\forall~\sigma\in {\bf S}_{n}$.
\end{cor}

\section{Some Properties of $\cal K_{[n,\kappa]}$}
In this section, we mainly discuss some properties of $\cal K_{[n,\kappa]}$ via its structure vector, which reveal the inside structure of skew-symmetric games. Particularly, they will be used in the sequel for investigating the decomposition of finite games.

\subsection{Existence of ${\cal K}_{[n,\kappa]}$}

The following proposition shows that when $n>\kappa+1$ the ${\cal K}_{[n,\kappa]}$ does not exit.

\begin{prp}\label{p4.1}Consider $G\in {\cal K}_{[n;\kappa]}.$ If $n>\kappa+1$, then
\begin{align}\label{4.40}
c_{i}(x)=0,\quad \forall~x\in S,~i=1,2,\cdots,n.
\end{align}
\end{prp}
\noindent{\it Proof.} Assume $n>\kappa+1$. It is easy to  know that for each $x=(x_{1},x_{2},\cdots,x_{n})\in S,$ there exists at least two strategies $x_{i}, x_{j}$ ($i>1$ and $j>1$), satisfying
\begin{align}\label{4.2}
x_{i}=x_{j},~i\neq j.
\end{align}
Let $\sigma=(i,j)$. Applying (\ref{4.2}) to (\ref{2.2}), we have
$$
\begin{array}{l}
c_{1}(x_{1},\cdots,x_{i},\cdots,x_{j},\cdots,x_{n})=-c_{1}(x_{1},\cdots,x_{j},\cdots,x_{i},\cdots,x_{n})=0.
\end{array}
$$
Hence $$V_{1}=0.$$
Using (\ref{3.201}), we have
$$V_{i}=0,\quad i=2,3,\cdots,n.$$
The conclusion follows.\hfill $\Box$

Next, we consider the marginal case when $n=\kappa+1$.
\begin{prp}Consider $G\in {\cal K}_{[n;\kappa]}.$ If $n=\kappa+1$, then $G$ is a zero-sum game. That is, $$\sum\limits_{i=1}^{n}c_{i}x=0,~~\forall~x\in S.$$
\end{prp}

\noindent{\it Proof.}
Since $n=\kappa+1,$ for any $x=(x_{1},x_{2},\cdots,x_{n})\in S,$ there exists at least two strategies $x_{i},x_{j}$ satisfying $x_{i}=x_{j}.$
Let $\sigma=(i,j)$.  From (\ref{2.2}) it is easy to verify that
\begin{align}\label{4.9}
c_{p}(x)=\left\{
  \begin{array}{ll}
    0,~~~~~~~~p\neq i,j; \\
    -c_{i}(x),~p=j;\\
    -c_{j}(x),~p=i.
  \end{array}
\right.
\end{align}
That is, $$c_{i}(x)=-c_{j}(x).$$
Hence,
\begin{eqnarray}
~~~~~~~~~~~~~~~~~\sum\limits_{p=1}^{n}c_{p}(x)&=&\sum\limits_{p\neq i,j}c_{p}(x)+c_{i}(x)+c_{j}(x)\nonumber\\
&=&c_{i}(x)+c_{j}(x)=0.\nonumber
\end{eqnarray}
\hfill $\Box$
\subsection{The Basis of ${\cal K}_{[n;\kappa]}$}

In this subsection we construct a basis for ${\cal K}_{[n;\kappa]}.$ According to Proposition \ref{p4.1}, we only need to consider the case when $n\leq \kappa+1$. Otherwise, ${\cal K}_{[n;\kappa]}=\{0\}$.

Moreover, according to Proposition \ref{p3.1.1}, to get a basis of ${\cal K}_{[n;\kappa]},$ it is enough to find a basis for $V^c_1.$

\textbf{Algorithm 1:}

Define
$$
\begin{array}{ccl}
{\cal O}&=&\left\{s_{-1}=z=(z_1,\cdots,z_{n-1})\right.\left.\;\big|\; z_{1}<z_{2}\cdots<z_{n-1},z_{j}\in S_{j+1}\right\},
\end{array}
$$
where $s_{-1}\in S_{-1}.$

It is easy to see $|{\cal O}|=\ell$, where
\begin{align}\label{4.12}
\ell=\binom{\kappa}{n-1}=\frac{\kappa !}{(n-1)!(\kappa-n+1)!}.
\end{align}

Define a relation $\prec$ on the set ${\cal O}$ as follows:
\begin{align}\label{o5.2}
(z^{1}_{1},z^{1}_{2},\cdots,z^{1}_{n-1})\prec(z^{2}_{1},z^{2}_{2},\cdots,z^{2}_{n-1}),
\end{align}
if and only if, there exists $0< j\leq n-2$, such that
$$
\begin{cases}
z^{1}_{i}=z^{2}_{i},\quad 1\leq i\leq j\\
z^{1}_{j+1}<z^{2}_{j+1}.
\end{cases}
$$
One sees easily that the relation $\prec$ is a strict order which makes $\cal {O}$ a well ordered set as
$$
{\cal O}:=(z^1,z^2,\cdots,z^{\ell}).
$$
Now for each $z^i=(z^i_1,z^i_2,\cdots,z^i_{n-1})$, we define a set
\begin{align}\label{6.2.2}
{\cal O}_i:=\left\{z^i_{\sigma}=\left(z^i_{\sigma(1)}, \cdots,z^i_{\sigma(n-1)}\right)\;|\; \sigma\in {\bf S}_{n-1} \right\}.
\end{align}
According to the definition of skew-symmetry and the above construction, the following facts are either obvious or easily verifiable.
\begin{enumerate}
\item[Fact~~1:] If $z\not\in \bigcup_{i=1}^{\ell} {\cal O}_i$, then for any $x_1\in S_1$, $c_1(x_1, z)=0$, because at least two components of $z$ are the same.

\vspace{2mm}
\item[Fact~~2:]\begin{align}\label{4.13}
c_1(x_1, z^i_{\sigma})=\sign(\sigma)c_1(x_1,z^i),\quad x_1\in S_1;
\end{align}

\vspace{2mm}
\item[Fact~~3:]
$$
{\cal O}_i\bigcap {\cal O}_j=\emptyset,\quad i\neq j.
$$
\end{enumerate}

Now, it is clear that to construct the basis for $V^c_1$, we need only to construct a ``dual basis" for each ${\cal O}_i$. Note that $x_1\in S_1$ is free, then we can define
\begin{equation}\label{4.20}
\begin{array}{lll}
\left(\eta^i_j\right)^T:=\d_{\kappa}^j\dsum_{\sigma\in {\bf S}_{n-1} }\sign(\sigma)\d_{\kappa}^{z^i_{\sigma(1)}} \d_{\kappa}^{z^i_{\sigma(2)}}\cdots \d_{\kappa}^{z^i_{\sigma(n-1)}},
\end{array}
\end{equation}
where $j=1,\cdots,\kappa,;~i=1,\cdots,\ell.$

Summarizing the above construction and argument yields the following results.

\begin{lem}\label{l4.5} Consider $G\in{\cal K}_{[n;\kappa]}$. Define
\begin{align}\label{4.14}
B=\begin{bmatrix}
\eta^1_1\\
\cdot\\
\eta^1_{\kappa}\\
\vdots\\
\eta^{\ell}_1\\
\cdot\\
\eta^{\ell}_{\kappa}\\
\end{bmatrix},
\end{align}
where $\eta^i_j$, $j=1,\cdots,\kappa$, $i=1,\cdots,\ell,$ are defined in (\ref{4.20}). Then, there exists a row vector $v\in \R^{\kappa\ell}$, such that
$$V_{1}^{c}=vB.
$$
\end{lem}
\begin{lem}\label{l4.6}
$$\left<\eta^{i_{1}}_{j_{1}},\eta^{i_{2}}_{j_{2}}\right>=\left\{
  \begin{array}{ll}
    0,~~~~~~~~~~~~(i_{1},j_{1})\neq(i_{2},j_{2}); \\
    (n-1)!,~~~~(i_{1},j_{1})=(i_{2},j_{2}),
  \end{array}
\right.$$
where $1\leq i_{1},i_{2}\leq\ell,~1\leq j_{1},j_{2}\leq \kappa.$
\end{lem}

According to Lemmas \ref{l4.5} and \ref{l4.6}, one sees that $\Row(B)$ is an orthonormal basis of $V^c_1$. Furthermore,   Using Proposition \ref{p3.1.1}, we have the following result:

\begin{thm}\label{t4.6} The basis of ${\cal K}_{[n;\kappa]}$ is $\Row(D)$, where
\begin{align}\label{basis.1}
D=\left[
B,-BW_{[\kappa^{0},\kappa]}W_{[\kappa,\kappa]},-BW_{[\kappa^{1},\kappa]}W_{[\kappa,\kappa^2]},\cdots,-BW_{[\kappa^{n-2},\kappa]}W_{[\kappa,\kappa^{n-1}]}\right].
\end{align}
\end{thm}

\begin{cor}\label{c4.7}
\begin{align}\label{c4.17}
\dim\left( {\cal K}_{[n;\kappa]}\right)=\kappa\ell:=\beta.
\end{align}
\end{cor}
Recall Examples \ref{e2.3} and \ref{e3.4}, where the dimensions of ${\cal K}_{[3;2]}$ and ${\cal K}_{[3;3]}$ have been proved to be $2$ and $9$ respectively. Both two dimensions verify the formula (\ref{c4.17}).

\begin{prp}\label{p4.7}
Define
$$
d_i=\Row_i(D),\quad i=1,\cdots,\beta.
$$
Then we have
\begin{align}\label{4.8}
\left<d_{i},d_{j}\right>=\left\{
  \begin{array}{ll}
    0,~~~~~~i\neq j; \\
    n!,~~~~~i=j.
  \end{array}
\right.
\end{align}
\end{prp}

\noindent{\it Proof.}~Denote$$C_{m}=W_{[\kappa^{m-2},\kappa]}W_{[\kappa,\kappa^{m-1}]},\quad m=2,3,\cdots,n.$$ From Theorem \ref{t4.6} we know
\begin{eqnarray}
d_{s}&=&Row_{s}(\Phi)\nonumber\\
&=&[Row_{s}(B),Row_{s}(-BC_{2}),\cdots,Row_{s}(-BC_{n})].\nonumber
\end{eqnarray}
According to Proposition \ref{pa.7}, it is easy to calculate that
$$\begin{array}{ll}
~~~~\left< d_{i},d_{j}\right>=d_{i} d^{T}_{j}\\
=Row_{i}(B)Row^{T}_{j}(B)+\sum\limits_{m=2}^{n}Row_{i}(-BC_{m})Row^{T}_{j}(-BC_{m})\\
=Row_{i}(B)Row^{T}_{j}(B)+\sum\limits_{m=2}^{n}Row_{i}(-B)C_{m}C^{T}_{m}Row^{T}_{j}(-B)\\
=n Row_{i}(B)Row^{T}_{j}(B).
\end{array}$$
(\ref{4.8}) follows from Lemma \ref{l4.6}.
\hfill $\Box$
\section{ Orthogonality of ${\cal K}_{[n;\kappa]}$ with ${\cal S}_{[n;\kappa]}$}

\subsection{Basis of ${\cal S}_{[n;\kappa]}$}
\begin{thm}\cite{che17}\label{t6.1} Consider $G\in {\cal S}_{[n;\kappa]}.$ Then
\begin{align}\label{t123}
dim({\cal S}_{[n;\kappa]})=\kappa p:=\alpha,
\end{align}
where$~p:=\binom{n+\kappa-2}{n-1}=\frac{(n+\kappa-2) !}{(n-1)!(\kappa-1)!}.$
\end{thm}

\begin{thm}\cite{che17}\label{t6.2}~$G\in {\cal G}_{[n;\kappa]}$ is a symmetric game, if and only if,
\begin{enumerate}
\item[(1)]
$V^c_1\left[I_{\kappa}\otimes \left(W_{[\kappa^{s-2}, \kappa]}W_{[\kappa, \kappa^{s-1}]}\right) -I_{k^{s+1}}\right]=0,\;~~s=2,3,\cdots,n-1.$

\vspace{2mm}
\item[(2)]
$V^c_i=V^c_1W_{[\kappa^{i-1}, \kappa]},\quad i=2,3,\cdots,n.$
\end{enumerate}
\end{thm}
For symmetric games, similar to Algorithm 1, we give an algorithm  to construct a basis for $V_{1}^{c}$.

\textbf{Algorithm 2:}

Define
$$\begin{array}{ccl}
{\cal Q}&=&\left\{s_{-1}=z=(z_1,\cdots,z_{n-1})\right.\left.\;\big|\; z_{1}\leq z_{2}\cdots\leq z_{n-1},z_{j}\in S_{j+1}\right\}.
\end{array}$$
Applying the relation $\prec$ defined in (\ref{o5.2}) to ${\cal Q}$, it also makes ${\cal Q}$ a well ordered set as
$$
{\cal Q}:=(z^1,z^2,\cdots,z^{p}).
$$
Now for each $z^i=(z^i_1,z^i_2,\cdots,z^i_{n-1})$, we define a set
\begin{eqnarray}\label{6.2.1}
{\cal Q}_i&:=&\left\{z^i_{\sigma}=\left(z^i_{\sigma(1)}, z^i_{\sigma(2)},\cdots,z^i_{\sigma(n-1)}\right)\;|\; \sigma\in {\bf S}_{n-1} \right\}\nonumber\\
&:=&\{z^{i,1}<z^{i,2}<\cdots<z^{i,q_{i}}\},
\end{eqnarray}
where $q_{i}$ is the number of different $z^i_{\sigma}$ in ${\cal Q}_i$. Denote the number of $j$ in $z^i$ as
$\#_i(j)$, $j=1,\cdots,\kappa$, $i=1,\cdots,p$. Then we have
\begin{align}\label{6.2.101}
q_i=\frac{\kappa !}{\prod_{j=1}^{\kappa}(\#_i(j))!},\quad i=1,\cdots,p.
\end{align}
Moreover, they also satisfy that
\begin{align}\label{0013}
\sum\limits_{i=1}^{p}q_{i}=\kappa^{n-1.}
\end{align}

Define
\begin{align}\label{6.3.2}
\begin{array}{ccl}
\left(\zeta^i_j\right)^T&:=&\d_{\kappa}^j\dsum_{t=1}^{q_{i}}\d_{\kappa}^{z^{i,t}_{1}} \d_{\kappa}^{z^{i,t}_{2}}\cdots \d_{\kappa}^{z^{i,t}_{n-1}}.\\
\end{array}
\end{align}
Summarizing the above construction and the argument, we can obtain the following lemmas.
\begin{lem}\label{l6.4} Consider $G\in{\cal S}_{[n;\kappa]}$. Define
\begin{align}\label{6.5}
H=\begin{bmatrix}
\zeta^1_1\\
\cdot\\
\zeta^1_{\kappa}\\
\vdots\\
\zeta^{p}_1\\
\cdot\\
\zeta^{p}_{\kappa}\\
\end{bmatrix},
\end{align}
where $\zeta^i_j$, $j=1,\cdots,\kappa$, $i=1,\cdots,p,$ are defined in (\ref{6.3.2}). Then, there exists a row vector $v\in \R^{\alpha}$, such that
$$V_{1}^{c}=vH.$$
\end{lem}

\begin{lem}\label{l6.6}
$$\left<\zeta^{i_{1}}_{j_{1}},\zeta^{i_{2}}_{j_{2}}\right>=\left\{
  \begin{array}{ll}
    0,~~~~~~(i_{1},j_{1})\neq(i_{2},j_{2}); \\
    q_{i},~~~~~(i_{1},j_{1})=(i_{2},j_{2}),
  \end{array}
  \right.
  $$
where $ 1\leq i_{1},i_{2}\leq p,~1\leq j_{1},j_{2}\leq \kappa.$
\end{lem}

According to Lemmas \ref{l6.4}, \ref{l6.6}, and Theorem \ref{t6.2}, we have the following result:

\begin{thm}\label{t6.7}
\begin{enumerate}
\item[(1)]
The basis of ${\cal S}_{[n;\kappa]}$ is $\Row(E)$, where
\begin{align}\label{basis.2}
E=\begin{bmatrix}
H,HW_{[\kappa,\kappa]},\cdots,
HW_{[\kappa^{n-1},\kappa]}\end{bmatrix}.
\end{align}

\item[(2)]
Define
$$
e_i=\Row_i(E),\quad i=1,\cdots,\alpha.
$$
Then we have
\begin{equation}\label{6.9}
\left<e_{i},e_{j}\right>=\left\{
  \begin{array}{ll}
    0,~~~~~~~~i\neq j; \\
    nq_{i},~~~~~i=j.
  \end{array}
  \right.
\end{equation}
\end{enumerate}
\end{thm}
\subsection{Orthogonality Between Symmetric and Skew-Symmetric Subspaces}
Putting the two basis-set matrices $D$ and $E$ together, we construct a new basis-set matrix $Q$ as follows:
\begin{equation}\label{6.14}
Q=\begin{bmatrix}
D\\
E
\end{bmatrix}.
\end{equation}
We will prove that $Q$ is of full row rank. It is enough to prove the following proposition, which shows that ${\cal S}_{[n;\kappa]}$ and ${\cal K}_{[n;\kappa]}$ are two orthogonal subspaces.

\begin{prp}\label{p6.10}
\begin{align}\label{6.11}
\left<d_{i},e_{j}\right>=0,\quad ~i=1,2,\cdots,\beta,~j=1,2,\cdots,\alpha.
\end{align}
\end{prp}

\noindent{\it Proof.}~To prove this proposition, it is enough to verify the following result: for $\eta^{j}_{t}$ and $\zeta^{i}_{s}$ constructed by using ${\cal O}_{j}$ and $\cal{Q}_{i}$ respectively, we have
\begin{equation}\label{9.1}
\begin{array}{ccl}
\eta^{j}_{t}(\zeta^{i}_{s})^{T}=0,
\end{array}
\end{equation}
where $t,s=1,2,\cdots,\kappa;~j=1,2,\cdots,\ell;~i=1,2,\cdots,p.$

According to the construction of ${\cal O}_{j}$ and ${\cal Q}_{i}$, it can be found that for each ${\cal O}_{j}$, there exists unique ${\cal Q}_{i(j)}$, such that
\begin{align}\label{8.0}
z^{j}=z^{i(j)},~{\cal O}_{j}={\cal Q}_{i(j)}.
\end{align}
In this case, from (\ref{4.20}) and (\ref{6.3.2}) we have
\begin{equation}
\begin{array}{ccl}
\left(\eta^j_{t}\right)^T&:=&\d_{\kappa}^{t}\dsum_{\sigma\in {\bf S}_{n-1}}\sign(\sigma)\d_{\kappa}^{z^j_{\sigma(1)}} \d_{\kappa}^{z^j_{\sigma(2)}}\cdots \d_{\kappa}^{z^j_{\sigma(n-1)}},\nonumber\\
\left(\zeta^{i(j)}_{s}\right)^T&:=&\d_{\kappa}^{s}\dsum_{\sigma\in {\bf S}_{n-1}}\d_{\kappa}^{z^{i(j)}_{\sigma(1)}} \d_{\kappa}^{z^{i(j)}_{\sigma(2)}}\cdots \d_{\kappa}^{z^{i(j)}_{\sigma(n-1)}},\nonumber
\end{array}
\end{equation}
where $t,s=1,2,\cdots,\kappa.$

Note that
$$\begin{array}{l}
~~~|\{\sigma\in {\bf S}_{n-1}|\sigma~is~odd~permutation\}|=|\{\sigma\in {\bf S}_{n-1}|\sigma~is~even~permutation \}|=\frac{(n-1)!}{2}.
\end{array}$$
Then, it is easy to calculate that
\begin{equation}\label{6.12}
\begin{array}{ccl}
\eta^{j}_{t}(\zeta^{i(j)}_{s})^{T}=0,\quad t,s=1,2,\cdots,\kappa,~j=1,2,\cdots,\ell.
\end{array}
\end{equation}
Next, for $i\neq i(j)$,  we have ${\cal O}_{j}\cap {\cal Q}_{i}=\emptyset,$ it is easy to see that
\begin{align}\label{6.13}
\eta^{j}_{t}(\zeta^{i}_{s})^{T}=0,~1\leq t,s\leq \kappa.
\end{align}
In fact, the location of nonzero elements, i.e., ``1" and ``-1," in $\eta^{j}_{t}$ is different from that in $\zeta^{i}_{s}.$ Hence (\ref{6.13}) holds.
(\ref{6.12}) and (\ref{6.13}) imply (\ref{9.1}), which proves (\ref{6.11}).
\hfill $\Box$

According to Proposition \ref{p4.7}, Theorem \ref{t6.7} and Proposition \ref{p6.10}, the following result can be obtained.
\begin{prp}\label{p6.15}Q has full row rank. That is,
$$dim(Q)=\beta+\alpha.$$
\end{prp}

Summarizing Proposition \ref{p4.1} and Proposition \ref{p6.15} yields the following proposition.
\begin{prp}\label{1444}
\begin{enumerate}
  \item if $n~>~\kappa+1,$ then
  \begin{align}
  {\cal G}_{[n;\kappa]}={\cal S}_{[n;\kappa]}\oplus {\cal E}_{[n;\kappa]};
  \end{align}
  \item if $n~\leq~\kappa+1,$ then
  \begin{align}
  {\cal G}_{[n;\kappa]}={\cal S}_{[n;\kappa]}\oplus {\cal K}_{[n;\kappa]}\oplus{\cal E}_{[n;\kappa]}.
  \end{align}
\end{enumerate}
\end{prp}

\section{Decomposition of a Finite Game}

Given $G\in {\cal G}_{[n;\kappa]}$, we would like to decompose it into three subgames: $G_{S}\in {\cal S}_{[n;\kappa]}$, $G_{K}\in {\cal K}_{[n;\kappa]}$ and $G_{E}\in {\cal E}_{[n;\kappa]}$.
Precisely, we want to decompose the structure vector of $G$, that is $V_G$, into three parts as
\begin{align}\label{8.5.1}
V_G=V_G^{S}+V_G^{K}+V_G^{E},
\end{align}
where $V_G^{S}$, $V_G^{K}$, and $V_G^{E}$ are the structure vectors of $G_{S}$, $G_{K}$, and $G_{E}$ respectively.

Now set
\begin{align}\label{8.5.3}
X=(X_1,X_2),
\end{align}
where $X_1\in \R^{\b}$, $X_2\in \R^{\a}$ (when  $n>\kappa+1$ we have $\b=0$).

Then we have the following decomposition:
\begin{prp}\label{p8.5.1} Assume $G\in {\cal G}_{[n;\kappa]}$ with its structure vector $V_G$. Then
$$
V_G=V_G^{S}\oplus V_G^{K}\oplus V_G^{E},
$$
where
\begin{equation}\label{8.5.4}
\begin{array}{ccl}
X&=&V_GQ^{\mathrm{T}}(QQ^{\mathrm{T}})^{-1},\\
V_G^{K}&=&X_1D,\\
V_G^{S}&=&X_2E,\\
V_G^{E}&=&V_G-V_G^{S}-V_G^{E},\\
\end{array}
\end{equation}
and $X_i$, $i=1,2,$ are defined in (\ref{8.5.3}), $Q$ is defined in (\ref{6.14}).
\end{prp}

\noindent{\it Proof.}~From Theorem \ref{t4.6} and Theorem \ref{t6.7} we have
\begin{equation}\label{1485}
V_G^{K}=X_{1}D,~~V_G^{S}=X_{2}E.
\end{equation}
Putting (\ref{1485}) into (\ref{8.5.1}) yields
\begin{equation}\label{1488}
\begin{array}{ccl}
V_G&=&V_G^{K}+V_G^{S}+V_G^{E}\\
   &=&X_{1}D+X_{2}E+V_G^{E}\\
   &=&XQ+V_G^{E}.
\end{array}
\end{equation}
According to Proposition \ref{p6.15} and Definition \ref{2.3}, we know that $QQ^{\mathrm{T}}$ is nonsingular and
\begin{equation}\label{1496}
V_G^{E}Q^{\mathrm{T}}=0.
\end{equation}
Applying (\ref{1496}) with (\ref{1488}), it can be calculated that
$$X=V_GQ^{\mathrm{T}}(QQ^{\mathrm{T}})^{-1}.$$
The conclusion follows.
\hfill $\Box$
%
\section{An Illustrative Example}

This section provides an example to demonstrate the overall symmetry-based decomposition process of a finite game.

\begin{exa}\label{e100.3} Consider ${\cal G}_{[3;2]}$.
The bases of ${\cal K}_{[3;2]}$ and ${\cal S}_{[3;2]}$ are constructed respectively as follows:

\begin{itemize}
\item[(1)] ${\cal K}_{[3;2]}$:

\vspace{2mm}
It is easy to calculate that $\ell=1$ and
\begin{equation}\label{e.1}
\begin{array}{ccl}
{\cal{O}_{1}}&:=&\{\cal{O}_{1}^{1}=\{z^{1}_{\sigma}\mid~sign(\sigma) =1\}=\{z^{1,1}=(12)\},\\
~&&~\cal{O}_{1}^{2}=\{z^{1}_{\sigma}\mid~sign(\sigma)=-1\}=\{z^{1,2}=(21)\}\}.\\
\end{array}
\end{equation}
Using (\ref{4.20}), we calculate that
$$
\begin{array}{lll}
B_1^T&:=&\dsum_{\sigma\in {\bf S}_2} sign(\sigma)\d_{\kappa}^{z^1_{\sigma(1)}}\d_{\kappa}^{z^1_{\sigma(2)}}\d_{\kappa}^{z^1_{\sigma(3)}}=\d_2^1\d_2^2-\d_2^2\d_2^1
\end{array}
$$
Then
$$\eta^1_1=B_1(\d_2^1)^T,\quad\eta^1_2=B_1(\d_2^2)^T.$$
Finally, we have
\begin{equation}\label{e.2}
B=\begin{bmatrix}
B_{1}~~{\bf 0}^T_{4}\\
{\bf 0}^T_{4}~~B_{1}
\end{bmatrix}=\begin{bmatrix}
0&1&-1&0&0&0&0&0\\
0&0&0&0&0&1&-1&0
\end{bmatrix}.
\end{equation}
Using (\ref{basis.1}), the basis of ${\cal K}_{[3;2]}$ is obtained.
\\
\item[(2)] ${\cal S}_{[3;2]}$:

\vspace{2mm}
It can be calculated that $p=3$ and
\begin{equation}
\begin{array}{ccl}
\cal{Q}_{1}&:=&\{z^{1,1}=(11)\};\nonumber\\
\cal{Q}_{2}&:=&\{z^{2,1}=(12),z^{2,2}=(21)\};\nonumber\\
\cal{Q}_{3}&:=&\{z^{3,1}=(22)\}.
\end{array}
\end{equation}
Correspondingly, we have
$$q_1=1,~q_2=2,~q_3=1.$$
According to each ${\cal Q}_i$, we can calculate $\zeta^i$ using (\ref{6.3.2}).
To calculate $\zeta^1$ we have
$$
H_1^T:=\dsum_{t=1}^{q_1}\d_{2}^{z^{1,t}_1}\d_2^{z^{1,t}_2}=\d_2^1\d_2^1=\d_{4}^1.
$$
Then
$$
\begin{array}{ccl}
\zeta^1&=&\begin{bmatrix}
\zeta^1_1\\
\zeta^1_2\\
\end{bmatrix}=\begin{bmatrix}
(\d_2^1)^TH_1\\
(\d_2^2)^TH_1\\
\end{bmatrix}=\begin{bmatrix}
H_1&{\bf 0}_{4}^T\\
{\bf 0}_{4}^T&H_1
\end{bmatrix}.
\end{array}
$$

Similarly, we have
$$
\zeta^i=\begin{bmatrix}
H_i&{\bf 0}_{4}^T\\
{\bf 0}_{4}^T&H_i
\end{bmatrix},\quad i=2,3,
$$
where
$$
H_i^T:=\dsum_{t=1}^{q_i}\d_{2}^{z^{i,t}_1}\d_2^{z^{i,t}_2}
$$
are calculated as follows:
$$
\begin{array}{ccl}
H_2^T&:=&\d_{2}^{1}\d_{2}^{2}+\d_{2}^{2}\d_{2}^{1}=\d_{4}^{2}+\d_{4}^{3},\\
H_3^T&:=&\d_{2}^{2}\d_{2}^{2}=\d_{4}^{1}.\\
\end{array}
$$
Putting $\zeta^i$, $i=1,2,3$ together, we have
\begin{align}\label{e.3}
H=\begin{bmatrix}
\zeta^1\\
\zeta^2\\
\zeta^{3}
\end{bmatrix}=\begin{bmatrix}
1&0&0&0&0&0&0&0\\
0&0&0&0&1&0&0&0\\
0&1&1&0&0&0&0&0\\
0&0&0&0&0&1&1&0\\
0&0&0&1&0&0&0&0\\
0&0&0&0&0&0&0&1
\end{bmatrix}.
\end{align}
Using (\ref{basis.2}), the basis of ${\cal S}_{[3;2]}$ is obtained.
\\
\item[(3)]A Numerical Example:

Assume $G\in  {\cal G}_{[3;2]}$ with $V^c_1=\d_{8}^3$, $V^c_2=\d_{8}^6$, $V^c_3=\d_{8}^{7}$. Now we give the orthogonal decomposition of $G$ into $G_{K}$, $G_{S}$, and $G_{E}.$
The basis of ${\cal S}_{[3,2]},$ ${\cal K}_{[3,2]}$ and ${\cal E}_{[3,2]}$ is respectively as below:
$$\begin{array}{ccl}
E&=&[H,HW_{[2,2]},HW_{[4,2]}],\\
D&=&[B,-BW_{[2,2]},-BW_{[2,2]}W_{[2,4]}],
\end{array}$$
and $\alpha=6,~\beta=2.$

\vspace{2mm}
According to Proposition \ref{p8.5.1}, we can calculate that
\begin{equation}
\begin{array}{lll}
X_{1}=[-0.1667,0]\in \R^{2};\nonumber\\
X_{2}=[0,0,0.1667,0,0.6667,0]\in \R^{6};\nonumber\\
\end{array}
\end{equation}
and
$$\begin{array}{ccl}
V_G&=&[(\d_{8}^{3})^{T},(\d_{8}^{6})^{T},(\d_{8}^{7})^{T}]\nonumber\\
&=&V_G^{K}+V_G^{S}+ V_G^{E},
\end{array}$$
where
\begin{equation}
\begin{array}{lll}
V_G^{K}=X_1D\nonumber\\
~~~~~=[0,-0.1667,0.1667,0,0,0,0,0,0,0.1667,0,0,-0.1667,0,0,0,0,0,-0.1667,0,0.1667,0,0,0],\nonumber\\
V_G^{S}=X_2E\nonumber\\
~~~~=[0,0.1667,0.6667,0,0,0,-0.1667,0,0,-0.3333,0,0,0,0.3333,0,0,0,0,0,0,-0.3333,0,0.3333,0],\nonumber\\
V_G^{E}=V_G-V_G^{K}-V_G^{S}\nonumber\\
~~~~~=[0,0,0.6666,-0.6666,0,0,0,0,0,-0.3333,0,0,0,0.3333,0,0,0,0,0,0,-0.3333,0,0.3333,0].\nonumber
\end{array}
\end{equation}
The corresponding payoff matrices are in Table \ref{Tab2593}-Table \ref{Tab2620} respectively.
Comparing Tables \ref{Tab2593}, \ref{Tab2607} with Example \ref{e2.3}, it is obvious that $G_{S}$ is symmetric and $G_{K}$ is skew-symmetric.
\begin{table}[!htbp] 
\centering \caption{Payoff matrix of $G_{S}$:\label{Tab2593}}
\scalebox{0.8}{\begin{tabular}{|c|c|c|c|c|c|c|c|c|}
\hline
$ c\backslash a$ & 111 & 112 & 121 & 122 & 211 & 212 & 221 & 222 \\
\hline
$c_1$ & 0 &  0.1667  &  0.1667  & 0.6667 & 0 & 0 & 0 &0\\
\hline
$c_2$ & 0 & 0.1667 & 0 & 0 & 0.1667 & 0.6667 & 0 & 0 \\
\hline
$c_3$ & 0 & 0 & 0.1667 & 0 & 0.1667 & 0 & 0.6667 & 0 \\
\hline
\end{tabular}}
\end{table}
\begin{table}[!htbp] 
\centering \caption{Payoff matrix of $G_{K}$:\label{Tab2607}}
\scalebox{0.8}{\begin{tabular}{|c|c|c|c|c|c|c|c|c|c|}
\hline
$ c\backslash a$ & 111 & 112 & 121 & 122 & 211 & 212 & 221 & 222 \\
\hline
$c_1$ & 0 & -0.1667  & 0.1667  & 0 & 0 & 0 & 0 & 0\\
\hline
$c_2$ & 0 & 0.1667  & 0 & 0 & -0.1667  & 0 & 0 & 0 \\
\hline
$c_3$ & 0 & 0& -0.1667  & 0 & 0.1667  & 0 & 0 & 0 \\
\hline
\end{tabular}}
\end{table}
\begin{table}[!htbp] 
\centering \caption{Payoff matrix of $G_{E}$:\label{Tab2620}}
\scalebox{0.8}{\begin{tabular}{|c|c|c|c|c|c|c|c|c|c|}
\hline
$ c\backslash a$ &111&112&121&122&211&212&221&222 \\
\hline
$c_1$ &0&0&0.6666&-0.6667&0&0&0&0\\
\hline
$c_2$ &0&-0.3334&0&0&0&0.3333&0&0\\
\hline
$c_3$ &0&0&0&0&-0.3334&0&0.3333&0\\
\hline
\end{tabular}}
\end{table}
\end{itemize}
\end{exa}

\section{Conclusion}
Skew-symmetric games are proposed and studied in this paper. First, for two player games, it is proved that the vector subspace of SSGs is the orthogonal complement of the subspace of symmetric games. Second, a necessary and sufficient condition is obtained to verify whether a finite game is skew-symmetric. Moreover, its linear representation is constructed to provide a convenient method for the verification. Third, two properties of ${\cal K}_{[n;\kappa]}$ are obtained. One is the existence of ${\cal K}_{[n;\kappa]}.$ It is shown that:
(i) if $n>\kappa+1,$ then ${\cal K}_{[n;\kappa]}=\{0\};$ (ii) if $n=\kappa+1,$ then $G\in{\cal K}_{[n;\kappa]}$ is a zero-sum game.
In addition, a basis of ${\cal K}_{[n;\kappa]}$ is constructed, which plays an important role in the decomposition of finite games. Then a basis of ${\cal S}_{[n;\kappa]}$ is also given and the orthogonality of ${\cal K}_{[n;\kappa]}$ with ${\cal S}_{[n;\kappa]}$ is proved.
Finally, the orthogonal decomposition of a finite game into symmetric, skew-symmetric and asymmetric subspaces is investigated. A decomposition formula is given and an illustrative example is presented.

\end{document}